\documentclass[conference]{IEEEtran}
\IEEEoverridecommandlockouts

\usepackage{cite}
\usepackage{amsmath,amssymb,amsfonts}
\usepackage{algorithmic}
\usepackage{algorithm}
\usepackage{graphicx}
\usepackage{textcomp}
\usepackage{xcolor}
\usepackage{booktabs}
\usepackage{multirow}
\usepackage{hyperref}
\usepackage{balance}
\usepackage{enumitem}
\usepackage{amsthm}
\usepackage{tikz}
\usetikzlibrary{positioning, shapes.geometric, arrows.meta}

\newcommand{\cmark}{\ding{51}}
\newcommand{\xmark}{\ding{55}}
\usepackage{pifont}

\def\BibTeX{{\rm B\kern-.05em{\sc i\kern-.025em b}\kern-.08em
    T\kern-.1667em\lower.7ex\hbox{E}\kern-.125emX}}

\begin{document}

\title{Evidential Trust-Aware Model Personalization in Decentralized Federated Learning for Wearable IoT}

\author{
\IEEEauthorblockN{Murtaza Rangwala, Richard O. Sinnott and Rajkumar Buyya}
\IEEEauthorblockA{
\textit{Quantum Cloud Computing and Distributed Systems (qCLOUDS) Lab}\\
\textit{School of Computing and Information Systems} \\
\textit{The University of Melbourne, Australia} \\
Email: \{rangwalam, rsinnott, rbuyya\}@unimelb.edu.au}
}


\maketitle

\begin{abstract}
Decentralized federated learning (DFL) enables collaborative model training across edge devices without centralized coordination, offering resilience against single points of failure. However, statistical heterogeneity arising from non-identically distributed local data creates a fundamental challenge: nodes must learn personalized models adapted to their local distributions while selectively collaborating with compatible peers. Existing approaches either enforce a single global model that fits no one well, or rely on heuristic peer selection mechanisms that cannot distinguish between peers with genuinely incompatible data distributions and those with valuable complementary knowledge. We present \textsc{Murmura}, a framework that leverages evidential deep learning to enable trust-aware model personalization in DFL. Our key insight is that epistemic uncertainty from Dirichlet-based evidential models directly indicates peer compatibility: high epistemic uncertainty when a peer's model evaluates local data reveals distributional mismatch, enabling nodes to exclude incompatible influence while maintaining personalized models through selective collaboration. \textsc{Murmura} introduces a trust-aware aggregation mechanism that computes peer compatibility scores through cross-evaluation on local validation samples and personalizes model aggregation based on evidential trust with adaptive thresholds. Evaluation on three wearable IoT datasets (UCI HAR, PAMAP2, PPG-DaLiA) demonstrates that \textsc{Murmura} reduces performance degradation from IID to non-IID conditions compared to baseline (0.9\% vs.\ 19.3\%), achieves 7.4$\times$ faster convergence, and maintains stable accuracy across hyperparameter choices. These results establish evidential uncertainty as a principled foundation for compatibility-aware personalization in decentralized heterogeneous environments.
\end{abstract}


\section{Introduction}
\label{sec:introduction}

Wearable Internet of Things (IoT) devices such as smartwatches, fitness trackers, and medical monitors continuously collect sensor data for applications ranging from human activity recognition to early disease detection~\cite{chen2020fedhealth}. Training machine learning models on such data traditionally requires centralized collection, raising privacy concerns and imposing substantial communication and compute overheads on bandwidth-constrained edge networks.

Federated Learning (FL) addresses these concerns by enabling collaborative model training while keeping data on user devices~\cite{mcmahan2017communication}. However, standard centralized FL approaches create a single point of failure and scalability bottleneck problematic for IoT deployments spanning thousands of heterogeneous devices. Decentralized Federated Learning (DFL) eliminates this bottleneck through peer-to-peer architectures where devices exchange and aggregate model updates directly~\cite{lalitha2018fully}, improving fault tolerance and reducing communication costs.

\subsection{The Personalization Challenge}

Each device's data follows a different distribution due to user diversity, sensor placement variations, environmental contexts, and device heterogeneity. This \textit{statistical heterogeneity} creates a fundamental tension: nodes need models personalized to their local distributions, yet naive personalization through isolated training wastes the potential benefits of collaboration.

The challenge is to enable each node to learn a model that performs well on its own data distribution while still benefiting from collaboration with peers that have related but not identical data. This requires each node to answer: \textit{which peers should I collaborate with?} A peer with very different data may hurt rather than help local model performance, even if that peer has trained effectively on their own distribution.

\subsection{The Peer Compatibility Problem}

Existing personalized FL methods typically assume centralized coordination where a server decides how to cluster or weight clients~\cite{tan2022towards}. In fully decentralized settings, each node must make local decisions about peer compatibility without central oversight.

The critical challenge is distinguishing between harmful dissimilarity and beneficial diversity. Peers may have genuinely incompatible data distributions, insufficient training quality, or complementary diversity that could enhance generalization. Simple similarity metrics based on model parameter distances or gradient norms cannot capture this distinction, as they conflate distributional differences with model quality. A peer with very different but high-quality data might be valuable for improving robustness, while a peer with similar but poorly-learned data might degrade performance.

\subsection{Our Approach: From Uncertainty to Trust}

To assess peer compatibility, nodes need more than just prediction accuracy; they need to understand \textit{why} a peer model makes its predictions. Consider evaluating a peer's model on local data: if the model predicts correctly, does that mean the peer has compatible data, or is the model simply guessing? If predictions are incorrect, does that indicate incompatibility, or is the data genuinely challenging?

Standard neural networks cannot answer these questions because they output only final predictions without indicating confidence or the type of uncertainty involved. This limitation makes it impossible to distinguish between a model that has never seen similar data (incompatible peer) versus one that finds the data inherently difficult (potentially valuable peer).

We address this by leveraging \textit{uncertainty quantification}. When a model evaluates data, we need to separate two fundamentally different types of uncertainty: uncertainty due to lack of knowledge about the data distribution (which signals incompatibility), and uncertainty due to inherent ambiguity in the data itself (which does not). Recent advances in evidential deep learning (EDL)~\cite{sensoy2018evidential} enable exactly this decomposition by having neural networks output not just predictions, but also quantified measures of these two uncertainty types.

We propose \textsc{Murmura}, a framework enabling \textit{trust-aware model personalization} in DFL through evidential uncertainty quantification. Each node evaluates candidate peer models by running them on local validation samples and examining both predictions and uncertainty characteristics. Peers whose models show low uncertainty from lack of knowledge receive high trust scores, indicating they were trained on compatible data distributions. Peers whose models show high uncertainty from lack of knowledge are excluded, as this reveals distributional mismatch. This approach naturally handles heterogeneity: each node personalizes by choosing which peers to aggregate based on compatibility, maintaining models adapted to local distributions while benefiting from selective collaboration.

\subsection{Contributions}

\noindent Our work makes the following contributions:

\begin{itemize}[leftmargin=*, labelsep=5.5pt]
    \item We introduce the first framework combining evidential uncertainty quantification with trust-aware personalization in DFL, enabling nodes to maintain personalized models through selective collaboration in non-IID environments.
    
    \item We develop an adaptive trust-aware aggregation algorithm that evaluates peer compatibility through cross-evaluation on local data and applies adaptive trust thresholds that tighten as training progresses.
    
    \item We provide \textsc{Murmura} as an open, extensible framework supporting multiple network topologies, baseline aggregation methods, and systematic evaluation under varying heterogeneity levels.
    
    \item Our approach computes uncertainty with a single forward pass per evaluation, avoiding the computational overhead of ensemble methods while providing richer information than point estimates.
\end{itemize}

\noindent The remainder of this paper is organized as follows. Section~\ref{sec:background} surveys related work. Section~\ref{sec:problem} formalizes the system model and design objectives. Section~\ref{sec:framework} presents the framework architecture. Section~\ref{sec:evidential} details the trust-aware personalization mechanism. Section~\ref{sec:implementation} describes implementation. Section~\ref{sec:evaluation} presents experimental evaluation. Section~\ref{sec:conclusion} concludes.

\section{Background and Related Work}
\label{sec:background}

This section surveys related work in DFL, personalization methods, and EDL, positioning \textsc{Murmura}'s contributions.

\subsection{Decentralized Federated Learning}

FL was introduced to enable collaborative model training while keeping data on local devices~\cite{mcmahan2017communication}. The original approach uses a central server to coordinate training rounds: clients train locally, send updates to the server, which aggregates them and broadcasts the result. While this protects data privacy, it creates centralization challenges.

Decentralized FL removes the need for a central server entirely. Instead, nodes form a network and exchange model updates directly with their neighbors. This peer-to-peer approach improves fault tolerance and can reduce communication costs by leveraging local network structure. Researchers have established both theoretical convergence guarantees~\cite{lalitha2018fully,koloskova2020unified} and practical benefits including faster training on bandwidth-limited networks~\cite{lian2017can}. For IoT networks where devices may join and leave dynamically, gossip-based communication provides particularly natural patterns for information exchange~\cite{hegedHus2019gossip}.

\subsection{Personalized Decentralized Federated Learning}

When devices have heterogeneous data distributions, forcing everyone to share a single global model often produces mediocre results. Personalized FL addresses this by letting each device maintain a model adapted to its own data while still benefiting from collaboration. Most personalized FL methods assume centralized control where a server decides how to group similar clients or weight their contributions~\cite{t2020personalized,li2021ditto}. 

In decentralized settings without a server, personalization becomes harder as nodes must autonomously decide which peers to collaborate with. Recent works have explored peer
selection mechanisms based on various similarity measures. BALANCE~\cite{fang2024byzantine} uses Euclidean distance between model parameters with adaptive thresholds to filter dissimilar peers, originally designed for Byzantine resilience but effectively enabling personalization through distance-based peer selection. Sketchguard~\cite{rangwala2025sketchguard} employs L2 distance on compressed model sketches to identify compatible peers while reducing communication costs. UBAR~\cite{guo2021byzantine} combines distance-based screening with loss-based evaluation in a two-stage filtering process. DisPFL~\cite{dai2022dispfl} and PFedDST~\cite{fan2025pfeddst} use gradient cosine similarity to assess peer compatibility for personalized aggregation. These approaches demonstrate that selective collaboration can improve personalization in decentralized settings.

\subsection{Evidential Deep Learning}
\label{sec:edl-background}

Standard neural network classifiers output probabilities for each class but cannot express confidence in these predictions. EDL~\cite{sensoy2018evidential} addresses this by having networks output evidence values that parameterize a Dirichlet distribution over probability vectors. This enables explicit quantification of epistemic uncertainty (lack of knowledge due to insufficient training or distribution shift) and aleatoric uncertainty (inherent data ambiguity). Epistemic uncertainty directly indicates whether a model has learned relevant patterns for given data, making it particularly useful for identifying compatible peers in heterogeneous FL settings. While EDL has been used in various FL contexts such as selecting reliable models~\cite{zhou2023trustworthy} or weighting aggregation in centralized settings~\cite{wang2023federated}, no prior work has leveraged it for compatibility-based peer selection in fully decentralized personalized FL under non-IID data conditions.

\subsection{Positioning and Research Gap}

Table~\ref{tab:comparison} compares \textsc{Murmura} with related approaches across key dimensions. Existing methods fall into two categories, each with fundamental limitations for our setting.

Decentralized personalized FL methods~\cite{fang2024byzantine, rangwala2025sketchguard, guo2021byzantine, dai2022dispfl, fan2025pfeddst} use heuristic similarity measures for peer selection. Distance-based methods treat all parameter differences equally regardless of whether they reflect distributional mismatch or legitimate model diversity. Gradient cosine similarity conflates training quality with distributional compatibility: two peers might have dissimilar gradients because one is well-trained on different data (incompatible but high-quality) or poorly trained on similar data (compatible but low-quality). These scenarios require opposite aggregation decisions but cannot be distinguished without understanding why models differ.

Uncertainty-aware aggregation methods face different limitations. Centralized approaches like~\cite{wang2023federated, t2020personalized, li2021ditto} rely on a server to evaluate uncertainty across all clients and make global decisions, unavailable when nodes operate autonomously. Prediction-level uncertainty methods cannot decompose epistemic (lack of knowledge) from aleatoric (inherent ambiguity) uncertainty. This decomposition is critical: a peer model showing high prediction uncertainty could indicate either incompatible training data or legitimately challenging samples. Without epistemic uncertainty quantification, nodes cannot distinguish these cases.

\textsc{Murmura} addresses both limitations by enabling autonomous compatibility assessment through epistemic uncertainty. High epistemic uncertainty during cross-evaluation directly reveals distributional mismatch regardless of prediction accuracy or gradient characteristics. The key novelty is applying evidential uncertainty decomposition to enable principled peer selection in fully decentralized personalized FL.

\begin{table}[t]
\centering
\caption{Comparison of \textsc{Murmura} with Related Work}
\label{tab:comparison}
\begin{tabular}{@{}lcccc@{}}
\toprule
\textbf{Approach} & \textbf{Decentral.} & \textbf{Personal.} & \textbf{Uncertainty} & \textbf{Compatibility} \\ 
\midrule
~\cite{t2020personalized} & \xmark & \cmark & \xmark & Central \\
~\cite{li2021ditto} & \xmark & \cmark & \xmark & Central \\
~\cite{fang2024byzantine} & \cmark & \cmark & \xmark & L2 distance \\
~\cite{rangwala2025sketchguard} & \cmark & \cmark & \xmark & L2 distance \\
~\cite{guo2021byzantine} & \cmark & \cmark & \xmark & Distance+Loss \\
~\cite{dai2022dispfl} & \cmark & \cmark & \xmark & Gradient sim. \\
~\cite{fan2025pfeddst} & \cmark & \cmark & \xmark & Gradient sim. \\
~\cite{wang2023federated} & \xmark & \xmark & \cmark & Central \\
\textbf{\textsc{Murmura}} & \cmark & \cmark & \cmark & Epistemic \\
\bottomrule
\end{tabular}
\end{table}

\section{Problem Formulation}
\label{sec:problem}

This section formalizes the system model and design objectives for trust-aware model personalization in DFL under statistical heterogeneity.

\subsection{System Model}

We consider $N$ nodes connected through an undirected graph $\mathcal{G} = (\mathcal{V}, \mathcal{E})$, where $\mathcal{V} = \{1, \ldots, N\}$ and $\mathcal{E}$ denote nodes and communication links respectively. Each node $i$ maintains a neighborhood $\mathcal{N}_i = \{j : (i,j) \in \mathcal{E}\}$ of directly reachable peers. Each node $i$ also possesses a local dataset $\mathcal{D}_i$ drawn from distribution $P_i(\mathbf{x}, y)$. We explicitly model statistical heterogeneity: $P_i \neq P_j$ in general. The learning objective is personalized models $\{\boldsymbol{\theta}_i\}_{i=1}^{N}$ minimizing aggregate risk:
\begin{equation}
    \min_{\{\boldsymbol{\theta}_i\}} \sum_{i=1}^{N} \mathbb{E}_{(\mathbf{x}, y) \sim P_i} [\ell(f(\mathbf{x}; \boldsymbol{\theta}_i), y)]
\end{equation}

\noindent where $f(\cdot; \boldsymbol{\theta})$ is the model and $\ell$ is the loss. Note each node optimizes for its own distribution $P_i$, enabling personalization while benefiting from selective collaboration. Training proceeds in synchronous rounds. In each round, nodes perform local training, exchange parameters with neighbors, and aggregate received updates according to their aggregation strategy.

\subsection{Statistical Heterogeneity Model}

We characterize the degree of non-IID-ness through distributional divergence between nodes. For nodes $i$ and $j$, we quantify their compatibility through:
\begin{equation}
    \Delta_{ij} = D_{\text{KL}}(P_i \| P_j)
\end{equation}
where $D_{\text{KL}}$ denotes Kullback-Leibler divergence. Nodes with $\Delta_{ij} \approx 0$ have similar distributions and should benefit from mutual collaboration. Nodes with large $\Delta_{ij}$ have incompatible distributions where naive aggregation would degrade personalized performance.

In practice, we do not have access to true distributions $P_i$ and cannot compute $\Delta_{ij}$ directly. Instead, nodes must infer compatibility from empirical observations—specifically, by evaluating how peer models perform on local data. This motivates our evidential approach where epistemic uncertainty serves as a proxy for distributional mismatch.

\subsection{Heterogeneity Sources in Wearable IoT}

For wearable sensor networks, statistical heterogeneity arises naturally from multiple sources:

\begin{itemize}[leftmargin=*, labelsep=5.5pt]
    \item \textbf{User diversity:} Different users exhibit distinct behavioral patterns.
    
    \item \textbf{Sensor placement:} Even for the same activity, sensor orientation and body placement create distributional variations.
    
    \item \textbf{Environmental context:} Indoor versus outdoor environments, terrain variations, and ambient conditions affect sensor readings.
    
    \item \textbf{Device heterogeneity:} Different sensor hardware, sampling rates, and calibration introduce systematic biases.
\end{itemize}

\noindent These factors create the non-IID conditions that necessitate personalization while making naive federated averaging suboptimal.

\subsection{Design Objectives}

\noindent\textsc{Murmura} is designed to satisfy four objectives:\\

\noindent \textbf{Objective 1: Personalized Learning.} Each node should learn a model that performs well on its own local data distribution. The model at node $i$ should minimize prediction error specifically on that node's data distribution $P_i$, rather than optimizing for a global average across all distributions.\\

\noindent \textbf{Objective 2: Selective Collaboration.} Nodes should benefit from collaboration with compatible peers while excluding incompatible peers whose data distributions are too dissimilar. Collaboration should improve performance over isolated training when compatible peers exist, while gracefully degrading to local learning when no compatible peers are available.\\

\noindent \textbf{Objective 3: Principled Compatibility Assessment.} Peer compatibility should be assessed through uncertainty quantification rather than heuristics or arbitrary thresholds. High epistemic uncertainty should indicate distributional incompatibility regardless of the source. High aleatoric uncertainty alone should not necessarily indicate incompatibility as it may reflect legitimately challenging data.\\

\noindent \textbf{Objective 4: Resource Efficiency.} The framework must be practical for deployment on resource-constrained wearable devices. Compatibility assessment should impose minimal computational overhead compared to training itself, and communication costs should not exceed the cost of exchanging model parameters.\\

\noindent The central challenge is enabling each node to make autonomous compatibility decisions in a fully decentralized setting, balancing personalization against collaboration while maintaining computational efficiency.

\section{Murmura Framework}
\label{sec:framework}

This section presents the \textsc{Murmura} framework, emphasizing the architectural decisions that enable evaluation of trust-aware personalization under controlled heterogeneity.

\subsection{Design Principles and Architecture}

Addressing the objectives defined in Section~\ref{sec:problem} requires an architecture that supports three critical capabilities: (1) isolation of aggregation strategy from network topology to enable controlled comparison, (2) reproducible heterogeneity simulation for systematic evaluation, and (3) extensibility for integrating new personalization approaches.

\begin{figure}[t]
\centering
\includegraphics[width=\columnwidth]{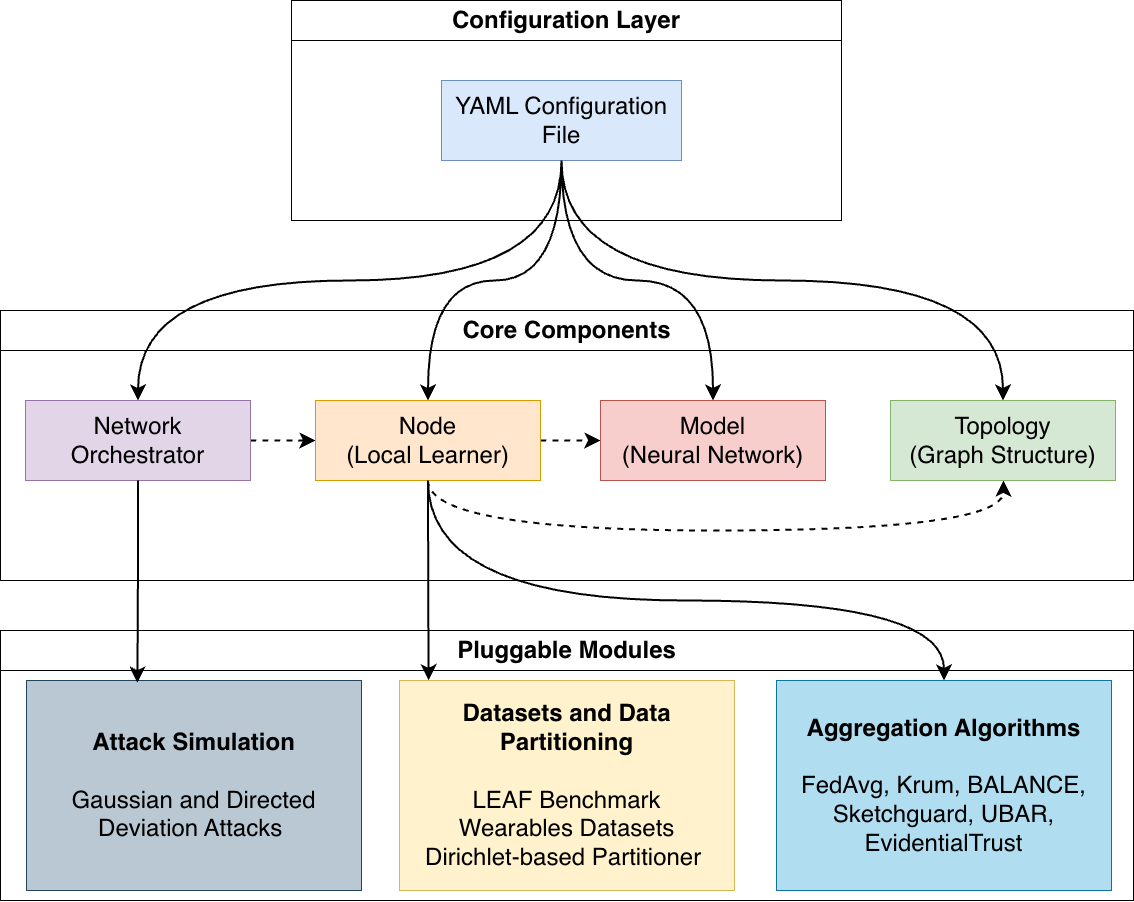}
\caption{Murmura framework architecture showing the three-layer design. The Configuration Layer provides YAML/JSON-based experiment specification. The Core Components layer contains the Network Orchestrator and Node, along with Topology and Model definitions. The Pluggable Modules layer enables flexible integration of aggregation strategies, data partitioning schemes and datasets.}
\label{fig:architecture}
\end{figure}

\textsc{Murmura} achieves these goals through modular separation of concerns, as illustrated in Figure~\ref{fig:architecture}. The architecture comprises three layers. The Core layer separates network orchestration (managing training rounds and communication) from node-level decision making (local training and aggregation). This separation is essential: it allows individual nodes to implement different aggregation strategies while maintaining identical communication patterns, isolating the impact of personalization mechanisms from topology effects.

The Pluggable Modules layer decouples topology generation from aggregation logic. This enables systematic study of how personalization strategies perform under different network structures (ring, fully-connected, Erdős-Rényi~\cite{erdos1960evolution}, k-regular) without conflating topology and algorithm effects—a common limitation in prior work where algorithms are evaluated on fixed topologies.

The Configuration layer addresses reproducibility through declarative experiment specification. Rather than hard-coding experimental parameters, researchers specify complete experimental configurations including topology, data partitioning, and hyperparameters in YAML files. This design choice supports two key research activities: systematic parameter sweeps (by generating configuration variants) and exact reproduction (by version-controlling complete experimental specifications).

\subsection{Controlled Heterogeneity Simulation}

A central challenge in evaluating personalized FL is creating realistic yet controlled non-IID data distributions. \textsc{Murmura} employs Dirichlet-based partitioning parameterized by concentration $\alpha$ to simulate statistical heterogeneity. Low $\alpha$ values create extreme heterogeneity where nodes receive samples predominantly from 1--2 classes, while higher values approach IID conditions. This parameterization enables systematic investigation of how personalization strategies degrade under increasing heterogeneity, addressing Objective 2 from Section~\ref{sec:problem}. Critically, the framework supports natural partitioning strategies (e.g., by subject ID for wearable datasets) alongside synthetic Dirichlet partitioning. This dual capability allows validation that findings from controlled synthetic heterogeneity generalize to realistic distribution patterns.

\subsection{Decentralized Training Protocol}

Algorithm~\ref{alg:training} formalizes the training protocol. The key design decision is synchronous execution with parallel local training followed by parallel aggregation. While asynchronous protocols might better reflect real deployments, synchronous execution eliminates timing effects as a confounding variable, enabling cleaner isolation of personalization strategy impact.

Each round proceeds through local training, parameter exchange with neighbors, and trust-aware aggregation. The aggregation step is where personalization occurs: each node autonomously evaluates received models and selectively incorporates compatible updates. Upon completion, each node maintains a personalized model adapted to its local data distribution. The protocol's simplicity is deliberate: by standardizing communication and training patterns, we ensure that performance differences across aggregation strategies reflect algorithmic merit rather than implementation artifacts. This design directly supports Objective 3 (principled compatibility assessment) by providing a controlled environment for evaluation.

\begin{algorithm}[ht]
\caption{\textsc{Murmura} Decentralized Training}
\label{alg:training}
\begin{algorithmic}[1]
\REQUIRE Nodes $\{1, \ldots, N\}$ with local data $\{\mathcal{D}_i\}$
\REQUIRE Topology $\mathcal{G}$ with neighborhoods $\{\mathcal{N}_i\}$
\REQUIRE Rounds $T$, local epochs $E$
\FOR{round $t = 1$ to $T$}
    \FOR{each node $i \in \mathcal{V}$ \textbf{in parallel}}
        \STATE $\boldsymbol{\theta}_i^{t} \leftarrow \text{LocalTrain}(\boldsymbol{\theta}_i^{t-1}, \mathcal{D}_i, E, t)$
    \ENDFOR
    \FOR{each node $i \in \mathcal{V}$ \textbf{in parallel}}
        \STATE Collect $\{\boldsymbol{\theta}_j^{t}\}_{j \in \mathcal{N}_i}$ from neighbors
        \STATE $\boldsymbol{\theta}_i^{t} \leftarrow \text{TrustAwareAggregate}(i, \boldsymbol{\theta}_i^{t}, \{\boldsymbol{\theta}_j^{t}\}_{j \in \mathcal{N}_i}, t)$
    \ENDFOR
\ENDFOR
\STATE \textbf{Output:} Each node $i$ maintains personalized model $\boldsymbol{\theta}_i^{T}$
\end{algorithmic}
\end{algorithm}

\section{Evidential Trust-Aware Personalization}
\label{sec:evidential}

This section explains how \textsc{Murmura} uses EDL to assess peer compatibility and enable personalized model aggregation in heterogeneous data environments.

\subsection{Evidential Uncertainty Quantification}

For a $K$-class classification problem, evidential networks output evidence values $\mathbf{e} = [e_1, \ldots, e_K]$ via:
\begin{equation}
    e_k = \exp(z_k), \quad \alpha_k = e_k + 1
\end{equation}
where $z_k$ is the standard network logit for class $k$, and $\alpha_k$ are Dirichlet concentration parameters. The Dirichlet strength $S = \sum_{k=1}^{K} \alpha_k$ measures the total accumulated evidence and indicates overall prediction confidence. The evidential framework decomposes predictive uncertainty into two meaningful components. \\

\noindent \textbf{Epistemic uncertainty} (lack of knowledge) measures whether the model has learned patterns relevant to the data:
\begin{equation}
    u = \frac{K}{S}
\end{equation}
where $K$ is the number of classes and $S$ is the Dirichlet strength. This uncertainty ranges from 0 (high confidence) to 1 (complete ignorance). High epistemic uncertainty signals that the model has not seen similar data during training, indicating distributional mismatch, which is precisely the signal needed to identify incompatible peers.\\

\noindent \textbf{Aleatoric uncertainty} (data ambiguity) captures inherent difficulty via prediction entropy:
\begin{equation}
    H[\hat{\mathbf{p}}] = -\sum_{k=1}^{K} \frac{\alpha_k}{S} \log \frac{\alpha_k}{S}
\end{equation}
High aleatoric uncertainty indicates genuinely ambiguous inputs (e.g., transitional movements between activities) rather than distributional incompatibility.\\

\noindent This distinction is crucial for personalization: a well-trained peer model evaluated on incompatible data shows high epistemic uncertainty (revealing distributional mismatch), whereas the same model on compatible but challenging data shows low epistemic uncertainty despite high aleatoric uncertainty. This enables principled compatibility assessment.

\subsection{Training Evidential Models}

Evidential models require a specialized loss function that encourages accurate predictions while preventing the model from simply claiming high confidence everywhere. Following Sensoy et al.~\cite{sensoy2018evidential}, we use:
\begin{equation}
    \mathcal{L} = \sum_{k=1}^{K} (y_k - \hat{p}_k)^2 + \lambda_t \cdot \text{KL}[\text{Dir}(\tilde{\boldsymbol{\alpha}}) \| \text{Dir}(\mathbf{1})]
\end{equation}
where $\mathbf{y} = [y_1, \ldots, y_K]$ is the one-hot encoded true label, $\hat{p}_k = \alpha_k / S$ is the expected probability for class $k$, and $\tilde{\boldsymbol{\alpha}} = \mathbf{y} + (1 - \mathbf{y}) \odot \boldsymbol{\alpha}$ removes evidence from the true class (where $\odot$ denotes element-wise multiplication). The first term is a squared error measuring prediction accuracy, while the KL divergence term prevents accumulating evidence for incorrect classes. The coefficient $\lambda_t$ controls the strength of the KL regularization and follows a linear annealing schedule:
\begin{equation}
    \lambda_t = \lambda_{\max} \cdot \min\left(1, \frac{t}{T_{\text{anneal}}}\right)
\end{equation}
where $t$ is the current training round, $T_{\text{anneal}}$ is the number of rounds over which to anneal (typically set to half the total training rounds), and $\lambda_{\max}$ is the maximum regularization weight. This linear annealing allows initial learning of accurate predictions before regularizing uncertainty estimates, preventing premature commitment to low-confidence predictions.

\subsection{Computing Compatibility from Uncertainty}

When node $i$ receives a model from peer $j$, it evaluates compatibility by running the peer's model on its own validation data. The evaluation examines both prediction accuracy and uncertainty characteristics, with epistemic uncertainty serving as the primary compatibility signal. Algorithm~\ref{alg:trust} formalizes this process. For each validation sample, the peer model produces both a prediction and an epistemic uncertainty score. High epistemic uncertainty indicates the peer's model was not trained on similar data, revealing distributional incompatibility regardless of whether predictions happen to be correct. The trust score naturally handles heterogeneity: a well-trained peer with compatible data shows low epistemic uncertainty (high trust) even if there are occasional prediction errors. A peer with incompatible data shows high epistemic uncertainty (low trust) regardless of prediction accuracy, as the peer's model was trained on different distributional patterns.

\begin{algorithm}[t]
\caption{Computing Compatibility via Cross-Validation}
\label{alg:trust}
\begin{algorithmic}[1]
\REQUIRE Validation set $\mathcal{D}_i^{\text{val}}$, peer model $\boldsymbol{\theta}_j$
\REQUIRE Uncertainty threshold $\tau_u$, accuracy weight $w_a$
\STATE Evaluate peer model on validation samples
\STATE Compute mean epistemic uncertainty $\bar{u}$ and accuracy $a$
\STATE Base trust score: $s_{\text{base}} = (1 - \bar{u}) \cdot (w_a \cdot a + (1 - w_a))$
\IF{$\bar{u} > \tau_u$}
    \STATE Apply penalty: $s_{\text{final}} = s_{\text{base}} \cdot \exp(-(\bar{u} - \tau_u))$
\ELSE
    \STATE $s_{\text{final}} = s_{\text{base}}$
\ENDIF
\RETURN $s_{\text{final}}$
\end{algorithmic}
\end{algorithm}

\subsection{Adaptive Trust Thresholds}

Trust thresholds should adapt as training progresses. Early in training, all models exhibit high uncertainty simply because they have not converged. Applying strict thresholds would reject all peers, preventing collaboration. As training proceeds, models accumulate evidence on their respective distributions and uncertainty decreases for in-distribution data. Following principles from~\cite{rangwala2025sketchguard}, we implement adaptive tightening:
\begin{equation}
    \tau_{\min}^{(t)} = \tau_{\min}^{(0)} \cdot \left(1 - \gamma_{\tau} \cdot e^{-\kappa \cdot t/T}\right)
\end{equation}
where $t$ is the current round, $T$ is the total number of rounds, $\tau_{\min}^{(0)}$ is the initial minimum trust threshold, $\gamma_{\tau} \in [0, 1]$ controls the maximum tightening amount, and $\kappa > 0$ controls the tightening rate. This schedule starts lenient (accepting peers despite high uncertainty) and gradually tightens (requiring lower uncertainty), matching the expected reduction in epistemic uncertainty as models converge on their respective data distributions.

\subsection{Trust-Aware Aggregation}

Each node independently decides which peers to trust and how much weight to assign them. Algorithm~\ref{alg:aggregation} shows the complete procedure. If no peers pass the compatibility threshold, the node simply retains its local model, providing graceful degradation to isolated learning. This autonomous decision-making enables personalization through selective collaboration rather than forced consensus or complete isolation. Compatibility assessment requires one forward pass per validation sample per neighbor, yielding complexity $O(d \cdot |\mathcal{D}_i^{\text{val}}| \cdot C_{\text{forward}})$ for $d$ neighbors. The evidential output layer adds minimal overhead (typically under 1\% additional inference time compared to standard classification).

\begin{algorithm}[t]
\caption{Trust-Aware Model Aggregation}
\label{alg:aggregation}
\begin{algorithmic}[1]
\REQUIRE Node $i$, local model $\boldsymbol{\theta}_i$, neighbor models $\{\boldsymbol{\theta}_j\}_{j \in \mathcal{N}_i}$
\REQUIRE Validation set $\mathcal{D}_i^{\text{val}}$, self-weight $\omega$, round $t$
\STATE Compute adaptive threshold $\tau_{\min}^{(t)}$ using Equation (6)
\STATE Evaluate each neighbor using Algorithm~\ref{alg:trust}, obtaining trust scores $\{s_j\}$
\STATE Retain only peers with $s_j \geq \tau_{\min}^{(t)}$
\IF{no peers pass threshold}
    \RETURN $\boldsymbol{\theta}_i$ \COMMENT{Keep local model}
\ENDIF
\STATE Normalize trust scores: $w_j = s_j / \sum_{k} s_k$ for retained peers
\STATE Aggregate trusted peers: $\boldsymbol{\theta}_{\text{peers}} = \sum_{j} w_j \cdot \boldsymbol{\theta}_j$
\STATE Personalize: $\boldsymbol{\theta}_i^{\text{new}} = \omega \cdot \boldsymbol{\theta}_i + (1 - \omega) \cdot \boldsymbol{\theta}_{\text{peers}}$
\RETURN $\boldsymbol{\theta}_i^{\text{new}}$
\end{algorithmic}
\end{algorithm}

\section{Implementation}
\label{sec:implementation}

This section describes the \textsc{Murmura}\footnote{The complete implementation of \textsc{Murmura} is available as open-source software at \url{https://github.com/Cloudslab/murmura}.} implementation, detailing the software architecture, key components, and practical considerations for deployment on resource-constrained wearable devices.

\subsection{Software Architecture and Design Principles}

\textsc{Murmura} is implemented in Python 3.12 using PyTorch~\cite{paszke2019pytorch} as the deep learning backend. The framework adopts a modular, layered architecture that cleanly separates concerns and facilitates extensibility. The codebase organization reflects the conceptual model presented in Section~\ref{sec:framework}, with each major component residing in its own package.

The core layer (murmura/core/) implements fundamental abstractions including the Network orchestrator, Node implementation, and type definitions for model states and data partitions. The topology layer (murmura/topology/) provides graph generators for ring, fully-connected, Erdős-Rényi, and k-regular topologies, all implementing a common Topology interface. The aggregation layer (murmura/aggregation/) contains implementations of baseline aggregators alongside our evidential trust aggregator, all adhering to a unified Aggregator protocol. The data layer (murmura/data/) supplies dataset adapters and partitioning utilities supporting both standard benchmarks and wearable sensor datasets. Finally, the configuration layer (murmura/config/) defines Pydantic schemas for type-safe, validated experiment specifications.

This modular design enables systematic comparison of aggregation strategies under identical conditions, facilitates integration of new datasets and topologies, and supports reproducible experimentation through declarative configuration.

\subsection{Evidential Deep Learning Components}

The evidential modeling capability requires two key components: an evidential output layer and a specialized loss function. The evidential output layer replaces the standard softmax head with a module that produces Dirichlet concentration parameters. This layer consists of a linear transformation from feature space to logit space, followed by an exponential activation ensuring non-negative evidence values, and finally the addition of one to obtain valid Dirichlet parameters. This modification is remarkably lightweight, adding only a single linear layer and activation function beyond a standard classifier. Evidential models compute uncertainty in a single forward pass, avoiding the computational overhead of ensemble methods or Monte Carlo dropout while providing richer information than point estimates. The computational overhead is empirically less than 1\% additional inference time compared to softmax classification.

The evidential loss function combines two terms as described in Section~\ref{sec:evidential}. The MSE component encourages accurate expected probability predictions, while the KL divergence component regularizes evidence accumulation on incorrect classes. Critically, the KL weight $\lambda$ anneals from zero to its maximum value over the first half of training. This annealing schedule allows the model to first learn accurate predictions before regularizing uncertainty estimates, preventing premature commitment to low-confidence predictions. In FL contexts, we use rounds rather than epochs for annealing since nodes may perform variable numbers of local epochs.

\subsection{Trust-Aware Aggregator Implementation}

The evidential trust aggregator implements Algorithm~\ref{alg:aggregation} with practical enhancements for stability and efficiency. Trust scores are computed through cross-evaluation, with evaluation limited to 100 validation samples to bound computational cost per neighbor. For networks with high connectivity, this limitation ensures that compatibility assessment remains tractable while providing sufficient statistical confidence in uncertainty estimates. The adaptive trust threshold implements tightening as described in Section~\ref{sec:evidential} using the exponential decay schedule with configurable parameters $\gamma_{\tau}$ and $\kappa$ that control the tightening behavior. Two key parameters control the personalization behavior. First, the self-weight parameter $\omega$ in Algorithm~\ref{alg:aggregation} determines how much influence the aggregated peer models have versus the local model. Lower $\omega$ values ($\omega < 0.3$) enable aggressive collaboration in homogeneous settings where all nodes have similar distributions. Higher values ($\omega > 0.7$) preserve personalization by limiting peer influence in highly heterogeneous settings. Second, the accuracy weight $w_a$ in Algorithm~\ref{alg:trust} balances the importance of prediction accuracy versus uncertainty characteristics when computing trust scores for peer compatibility evaluation.

\subsection{Wearable Sensor Dataset Integration}

A key challenge in FL research is bridging the gap between standard benchmark datasets and realistic sensor data characteristics. Wearable datasets often require preprocessing (sliding window segmentation, feature extraction, signal synchronization) that, if performed differently across studies, confounds experimental comparisons. 

\textsc{Murmura} addresses this through dataset adapters that standardize preprocessing while exposing key parameters (window size, stride, feature sets) as configuration options. This design enables researchers to systematically study how preprocessing choices interact with personalization strategies.

The framework provides adapters for three established wearable datasets (UCI HAR~\cite{anguita2013public}, PAMAP2~\cite{reiss2012introducing}, PPG-DaLiA~\cite{reiss2019deep}), each supporting multiple partitioning strategies. Critically, adapters support both synthetic Dirichlet partitioning (for controlled heterogeneity) and natural partitioning by subject ID (for realistic distribution patterns). This dual capability enables validation that findings from controlled experiments generalize to deployment-like conditions where heterogeneity arises from genuine user diversity.

The concentration parameter $\alpha$ in Dirichlet partitioning governs heterogeneity severity: $\alpha = 0.1$ creates extreme non-IID conditions where each node receives samples from only 1--2 classes, $\alpha = 0.5$ produces moderate heterogeneity, and $\alpha = 1.0$ approaches IID. This parameterized control enables systematic investigation of personalization performance across the heterogeneity spectrum, directly supporting the evaluation objectives in Section~\ref{sec:evaluation}.

\subsection{Reproducibility and Experimental Infrastructure}

Reproducibility is ensured through comprehensive seed management and configuration capture. At experiment initialization, seeds are set deterministically for Python's built-in random number generator, NumPy's random state, PyTorch's CPU random number generator, and all CUDA random number generators if GPU execution is enabled. Additionally, PyTorch's cuDNN backend is configured for deterministic operation, trading a small performance penalty for perfect reproducibility across runs.

All experimental parameters are specified via YAML configuration files following the schema defined in Section~\ref{sec:framework}. This declarative approach enables version control of experimental configurations, facilitates parameter sweeps through configuration templates, and ensures complete reproducibility by capturing all relevant settings.

\section{Performance Evaluation}
\label{sec:evaluation}

This section presents experimental evaluation of \textsc{Murmura}, examining personalization performance under varying data heterogeneity, convergence behavior, and sensitivity to hyperparameter choices.

\subsection{Experimental Setup}

\subsubsection{Datasets}
We evaluate on three established wearable sensor datasets for human activity recognition:

\begin{itemize}[leftmargin=*, labelsep=5.5pt]
    \item \textbf{UCI HAR}~\cite{anguita2013public}: Smartphone accelerometer and gyroscope data from 30 subjects performing 6 activities (walking, walking upstairs, walking downstairs, sitting, standing, lying). Contains 10,299 samples with 561 extracted features.

    \item \textbf{PAMAP2}~\cite{reiss2012introducing}: IMU data from 9 subjects performing 12 activities including household and exercise tasks. We extract 40 features per sliding window from three body-worn sensors.

    \item \textbf{PPG-DaLiA}~\cite{reiss2019deep}: Photoplethysmography and accelerometer signals from 15 subjects under real-life conditions. We formulate activity classification from 8 activity types with time-frequency features.
\end{itemize}

\subsubsection{Baselines}
We compare against four representative aggregation strategies:

\begin{itemize}[leftmargin=*, labelsep=5.5pt]
    \item \textbf{FedAvg}~\cite{mcmahan2017communication}: Standard federated averaging without Byzantine resilience or personalization mechanisms.

    \item \textbf{BALANCE}~\cite{fang2024byzantine}: Distance-based filtering with adaptive thresholds for Byzantine resilience and personalization.

    \item \textbf{Sketchguard}~\cite{rangwala2025sketchguard}: Count-Sketch compression with L2 distance based filtering for communication-efficient DFL.

    \item \textbf{UBAR}~\cite{guo2021byzantine}: Two-stage filtering combining distance-based screening with loss-based evaluation.
\end{itemize}

\subsubsection{Implementation Details}
We deploy 30 nodes in a fully-connected topology, training for 30 rounds with 5 local epochs per round. Models use a three-layer fully-connected architecture with ReLU activations and an evidential output layer for \textsc{Murmura}. We use SGD with learning rate 0.01 and batch size 32. For evidential trust aggregation hyperparameters, we set: self-weight $\omega=0.5$ to balance personalization and collaboration, accuracy weight $w_a = 0.5$ to equally weight prediction correctness and uncertainty, initial trust threshold $\tau_{\min}^{(0)} = 0.3$, adaptive tightening parameters $\gamma_{\tau} = 0.5$ and $\kappa = 1.0$, and epistemic uncertainty threshold $\tau_u = 0.7$ to filter high-uncertainty peers. All experiments use 5 random seeds; we report mean results across nodes.

\subsection{Research Questions}

Our evaluation addresses three research questions:

\begin{enumerate}[label=\textbf{RQ\arabic*:}, leftmargin=*, labelsep=5.5pt]
    \item How does evidential trust-aware aggregation affect model personalization under varying degrees of data heterogeneity?
    \item Does uncertainty-based peer selection improve convergence speed compared to baseline strategies?
    \item How sensitive is the framework to hyperparameter choices?
\end{enumerate}

\subsection{RQ1: Personalization Under Heterogeneity}

Table~\ref{tab:heterogeneity} presents classification accuracy across datasets and heterogeneity levels. Under high heterogeneity ($\alpha = 0.1$), \textsc{Murmura} achieves the highest accuracy on UCI HAR (95.4\%) and PAMAP2 (98.8\%), demonstrating effective personalization when data distributions diverge substantially across nodes. PPG-DaLiA exhibits lower absolute accuracy across all methods; we discuss the dataset-specific challenges and their implications in Section~\ref{sec:discussion}. The low standard deviation (7.3\% on UCI HAR compared to 18.4\% for FedAvg) indicates consistent performance across nodes despite heterogeneous local distributions.

\begin{table}[t]
\centering
\caption{Classification Accuracy (\%) Under Varying Heterogeneity}
\label{tab:heterogeneity}
\begin{tabular}{@{}llccc@{}}
\toprule
\textbf{Algorithm} & \textbf{$\alpha$} & \textbf{UCI HAR} & \textbf{PAMAP2} & \textbf{PPG-DaLiA} \\
\midrule
\multirow{3}{*}{FedAvg}
    & 0.1 & 85.6 & 90.5 & 26.6 \\
    & 0.5 & 97.9 & 92.6 & 66.5 \\
    & 1.0 & 98.5 & 89.2 & 72.9 \\
\midrule
\multirow{3}{*}{BALANCE}
    & 0.1 & 75.3 & 98.4 & 52.3 \\
    & 0.5 & 98.9 & 95.8 & 72.2 \\
    & 1.0 & 99.2 & 84.0 & 67.6 \\
\midrule
\multirow{3}{*}{Sketchguard}
    & 0.1 & 94.4 & 97.9 & 54.2 \\
    & 0.5 & 98.8 & 95.6 & 69.3 \\
    & 1.0 & 99.0 & 92.1 & 67.6 \\
\midrule
\multirow{3}{*}{UBAR}
    & 0.1 & 85.6 & 87.5 & 75.9 \\
    & 0.5 & 96.3 & 94.0 & 75.1 \\
    & 1.0 & 97.4 & 86.3 & 73.2 \\
\midrule
\multirow{3}{*}{\textbf{Murmura}}
    & 0.1 & \textbf{95.4} & \textbf{98.8} & 63.9 \\
    & 0.5 & 98.2 & 97.4 & \textbf{78.8} \\
    & 1.0 & 98.4 & 90.0 & 72.6 \\
\bottomrule
\end{tabular}
\end{table}

\begin{figure}[t]
\centering
\includegraphics[width=\columnwidth]{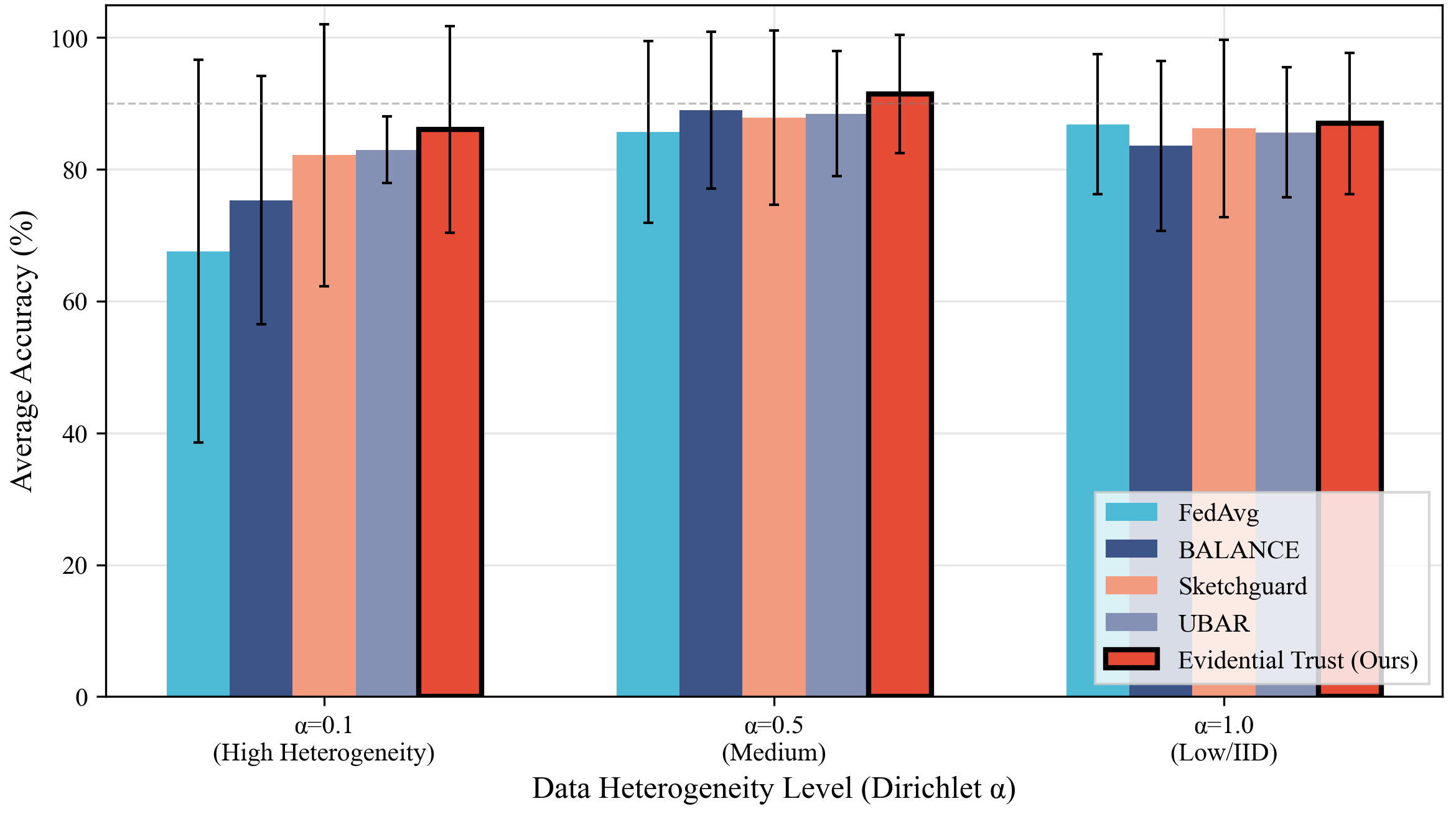}
\caption{Model accuracy across data heterogeneity levels (Dirichlet $\alpha$), averaged across all three datasets. Lower $\alpha$ indicates higher heterogeneity. \textsc{Murmura} (Evidential Trust) maintains consistent performance as heterogeneity increases.}
\label{fig:heterogeneity}
\end{figure}

Figure~\ref{fig:heterogeneity} visualizes performance across heterogeneity levels, averaged across all datasets. \textsc{Murmura} maintains stable accuracy as heterogeneity increases, while baseline methods exhibit more pronounced degradation. This stability stems from the trust-aware mechanism: under high heterogeneity, nodes with incompatible distributions exhibit elevated epistemic uncertainty during cross-evaluation, triggering exclusion from aggregation. Nodes effectively ``choose'' compatible peers, preserving personalized models.

\begin{figure}[t]
\centering
\includegraphics[width=\columnwidth]{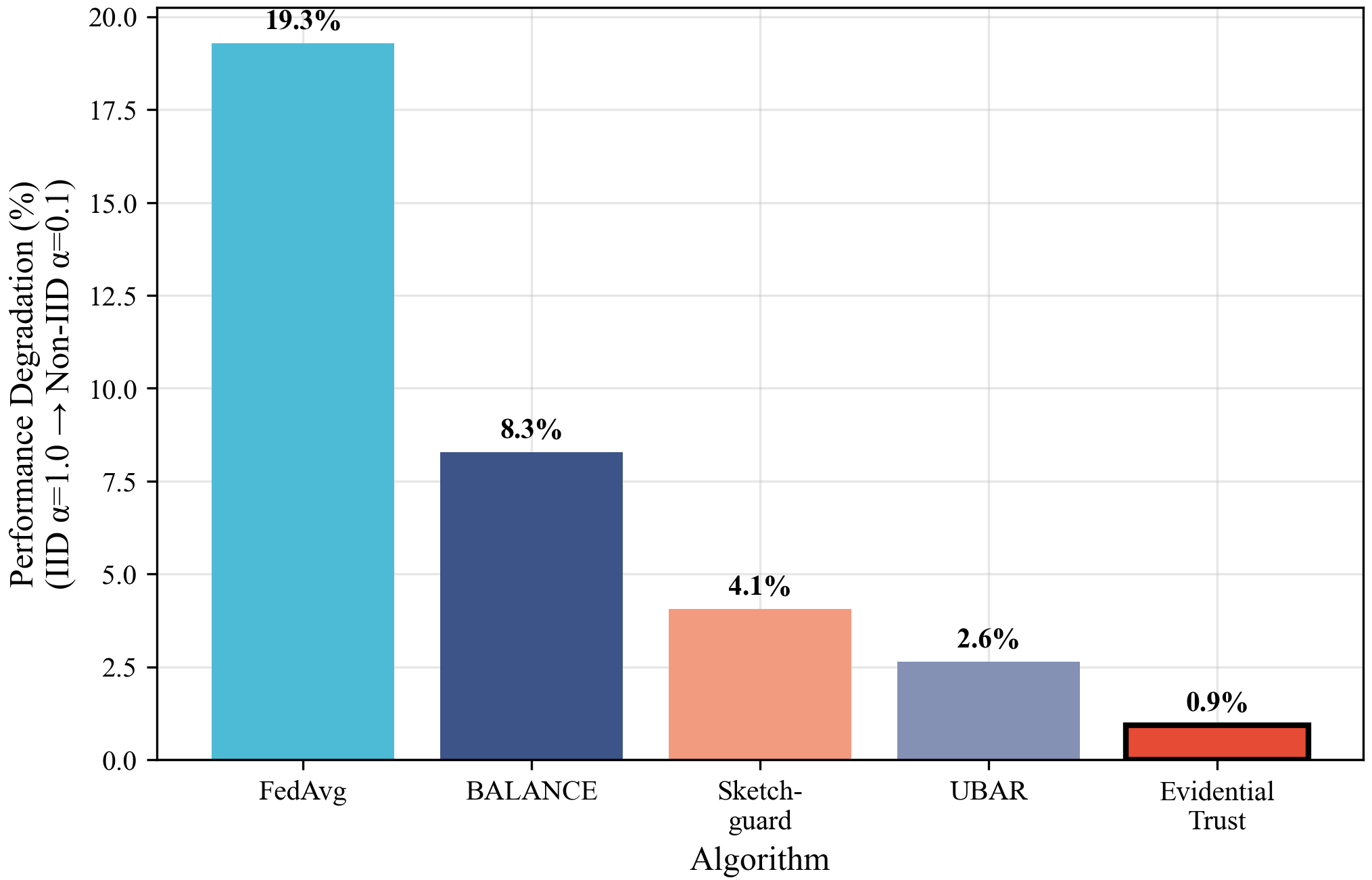}
\caption{Performance degradation from IID ($\alpha=1.0$) to non-IID ($\alpha=0.1$) conditions. Lower values indicate better robustness to heterogeneity. \textsc{Murmura} shows minimal degradation (0.9\%) compared to baselines.}
\label{fig:degradation}
\end{figure}

To quantify robustness to heterogeneity, we measure \textit{performance degradation}: the accuracy drop when transitioning from IID ($\alpha = 1.0$) to highly non-IID ($\alpha = 0.1$) conditions. Figure~\ref{fig:degradation} presents these results. \textsc{Murmura} exhibits the smallest degradation (0.9\% average across datasets), compared to FedAvg (19.3\%), BALANCE (8.3\%), Sketchguard (4.1\%), and UBAR (2.6\%). This 20$\times$ reduction in degradation compared to FedAvg demonstrates that evidential trust-aware aggregation effectively isolates nodes from incompatible peer influence.

\begin{figure}[t]
\centering
\includegraphics[width=\columnwidth]{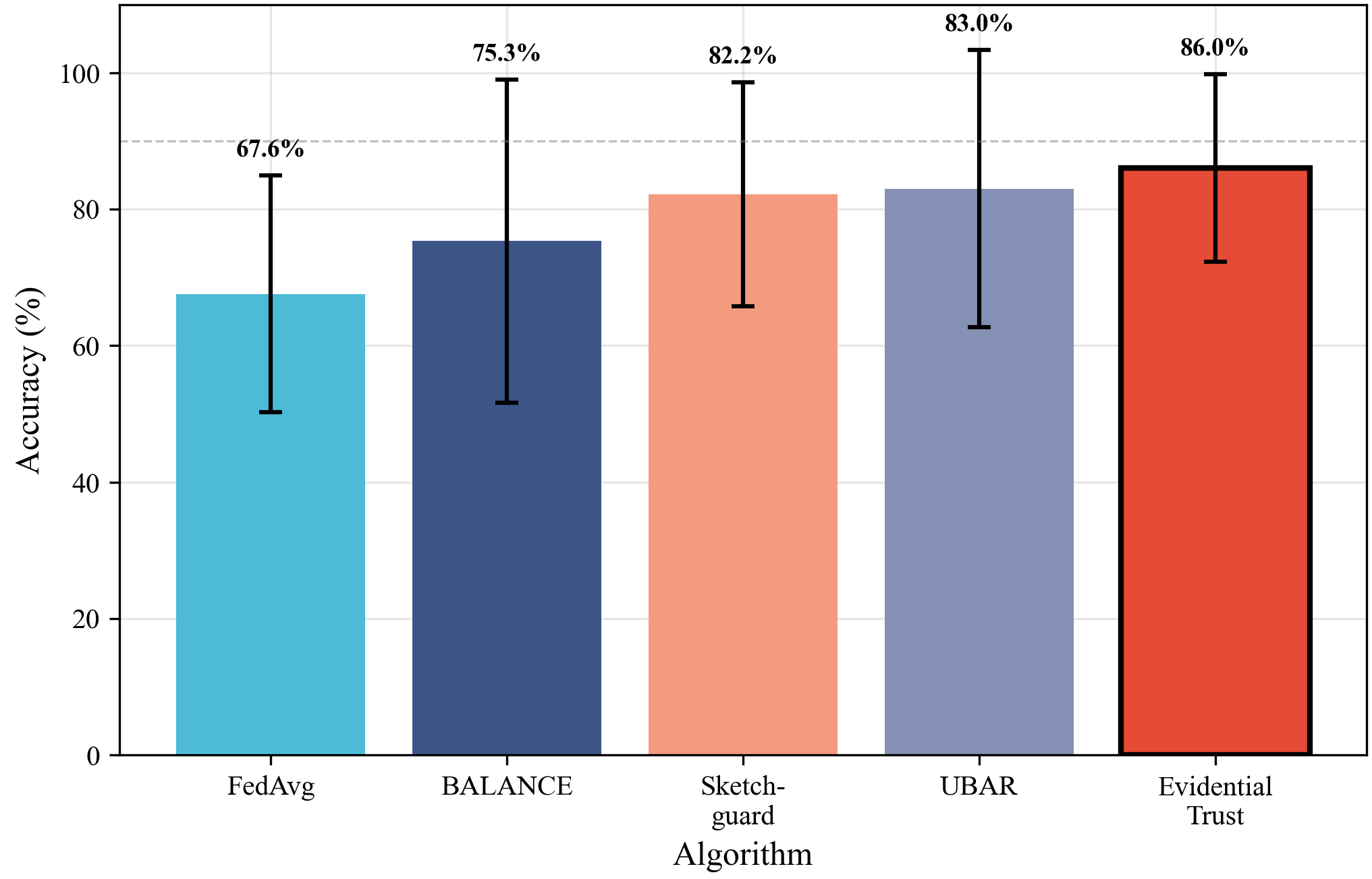}
\caption{Model personalization under high heterogeneity ($\alpha=0.1$). Bars show mean accuracy; error bars show standard deviation across nodes. \textsc{Murmura} achieves high accuracy with low variance across nodes.}
\label{fig:personalization}
\end{figure}

The personalization quality is further evidenced by examining per-node variance. Figure~\ref{fig:personalization} shows accuracy with standard deviation across nodes at $\alpha = 0.1$. \textsc{Murmura} achieves not only higher mean accuracy but also lower variance, indicating that personalization benefits all nodes rather than improving some at the expense of others.

\subsection{RQ2: Convergence Analysis}

\begin{figure}[t]
\centering
\includegraphics[width=\columnwidth]{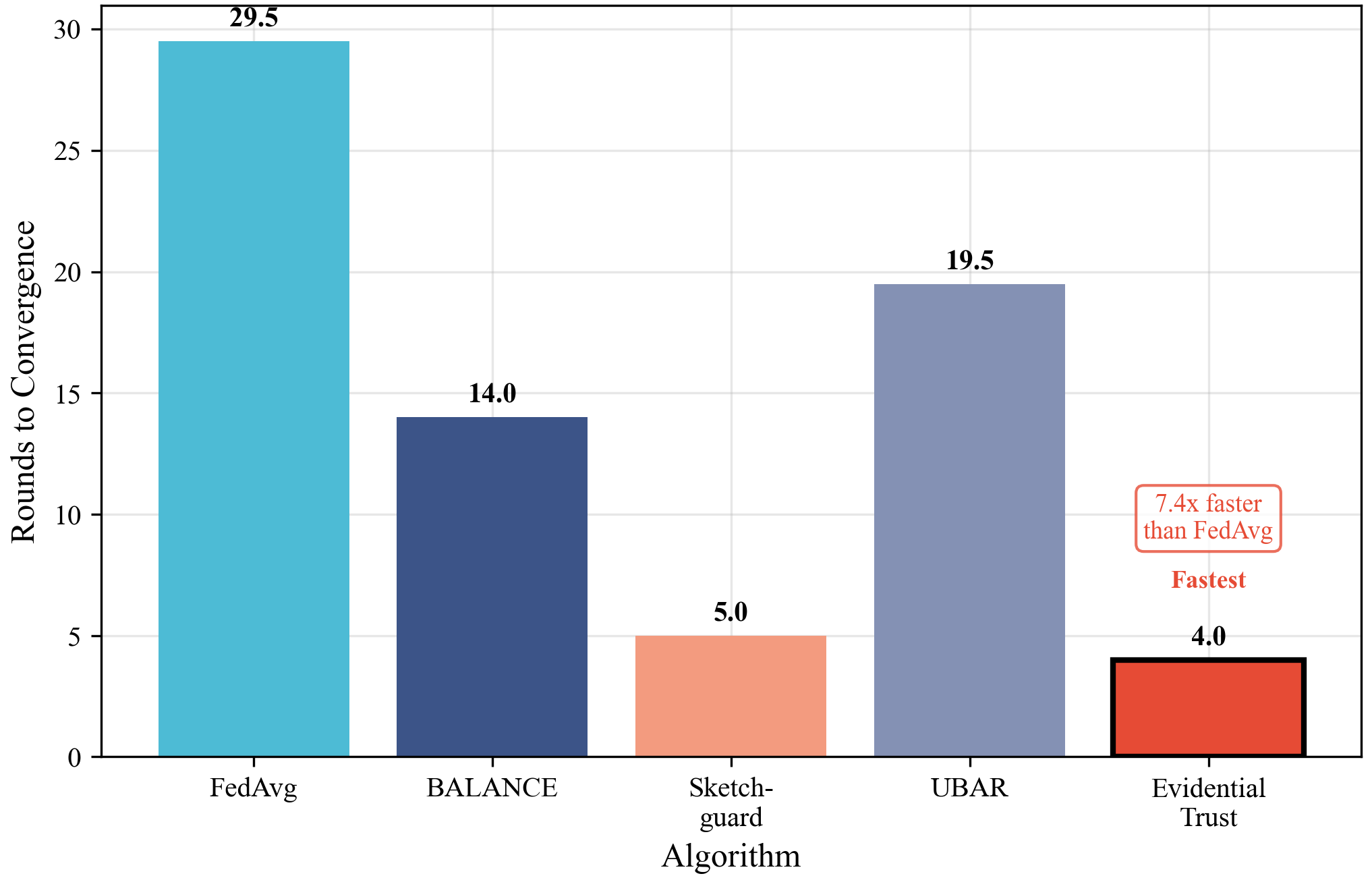}
\caption{Convergence speed comparison showing rounds to reach peak accuracy. \textsc{Murmura} converges 7.4$\times$ faster than FedAvg by filtering incompatible peer updates.}
\label{fig:convergence}
\end{figure}

Figure~\ref{fig:convergence} presents convergence speed measured as rounds required to reach peak accuracy. \textsc{Murmura} converges in 4.0 rounds on average, compared to 29.5 rounds for FedAvg, a 7.4$\times$ speedup. This acceleration arises from two factors: (1) evidential trust filtering excludes updates from incompatible or undertrained peers that would otherwise introduce noise, and (2) the personalized aggregation weights prioritize high-quality updates from compatible peers. The fast convergence has practical implications for wearable IoT deployments: fewer communication rounds reduce energy consumption and network bandwidth usage, critical constraints for battery-powered devices. Combined with the single-pass uncertainty computation, \textsc{Murmura} provides an efficient path to personalized models.

\subsection{RQ3: Hyperparameter Sensitivity}

Table~\ref{tab:ablation} presents ablation results examining sensitivity to key hyperparameters. For each parameter, we vary its value while holding others constant at defaults, reporting mean accuracy and standard deviation across the tested range.

\begin{table}[ht]
\centering
\caption{Ablation Study: Hyperparameter Sensitivity}
\label{tab:ablation}
\begin{tabular}{@{}lccc@{}}
\toprule
\textbf{Parameter} & \textbf{UCI HAR} & \textbf{PAMAP2} & \textbf{PPG-DaLiA} \\
\midrule
Accuracy weight ($w_a$) & 98.3 $\pm$ 0.3 & 97.1 $\pm$ 0.2 & 78.6 $\pm$ 0.4 \\
Self-weight ($\omega$) & 98.3 $\pm$ 0.5 & 96.8 $\pm$ 1.4 & 79.1 $\pm$ 3.7 \\
Trust threshold ($\tau_{\min}^{(0)}$) & 98.4 $\pm$ 0.2 & 97.3 $\pm$ 0.2 & 78.9 $\pm$ 0.2 \\
Uncertainty threshold ($\tau_u$) & 98.4 $\pm$ 0.2 & 97.2 $\pm$ 0.1 & 78.7 $\pm$ 0.1 \\
\bottomrule
\end{tabular}
\end{table}

The results demonstrate remarkable stability across hyperparameter choices. Standard deviations remain below 1\% for UCI HAR and PAMAP2 across all parameters, and below 4\% for the more challenging PPG-DaLiA dataset. The \textit{accuracy weight} ($w_a \in \{0.3, 0.5, 0.7, 0.9\}$) balances prediction accuracy against uncertainty in trust computation; performance remains stable across this range. The \textit{self-weight} ($\omega \in \{0.3, 0.5, 0.6, 0.7, 0.9\}$) controls personalization strength; higher values favor local models while lower values enable more collaboration. The \textit{trust threshold} ($\tau_{\min}^{(0)} \in \{0.05, 0.1, 0.2, 0.3\}$) and \textit{uncertainty threshold} ($\tau_u \in \{0.3, 0.5, 0.7, 0.9\}$) govern peer filtering stringency; the framework remains robust across a wide range. This stability reduces the need for extensive hyperparameter tuning in deployment, an important practical consideration for wearable IoT applications where per-device optimization is infeasible.

\subsection{Discussion}
\label{sec:discussion}

The experimental results support several conclusions regarding evidential trust-aware personalization:\\

\noindent \textbf{Personalization emerges from selective collaboration.} Rather than explicitly clustering nodes or learning separate models, \textsc{Murmura} achieves personalization through autonomous peer selection. Each node's model naturally adapts to its local distribution by incorporating updates only from compatible peers, while still benefiting from collaboration when compatible peers exist.\\

\noindent \textbf{Fast convergence reduces resource requirements.} The considerable speedup over baselines translates directly to reduced communication rounds, energy consumption, and time-to-deployment, all of which are critical factors for battery-constrained wearable devices.\\

\noindent \textbf{Robustness simplifies deployment.} The stability across hyperparameter choices reduces the burden of per-deployment tuning, making \textsc{Murmura} practical for heterogeneous IoT environments where individual device optimization is impractical.\\

\noindent The results on PPG-DaLiA, while showing improvement over baselines at moderate heterogeneity, reveal limitations under extreme heterogeneity. This dataset exhibits substantial inter-subject variability in physiological signals, creating scenarios where few truly compatible peers exist. In such cases, \textsc{Murmura}'s conservative filtering leads to near-isolated training, which, while preserving personalization, limits knowledge transfer. Future work could explore graduated trust mechanisms that extract value from moderately compatible peers.

\section{Conclusions and Future Work}
\label{sec:conclusion}

We presented \textsc{Murmura}, a framework enabling trust-aware model personalization in DFL through evidential uncertainty quantification. The key insight is that epistemic uncertainty directly indicates peer compatibility: high uncertainty when evaluating peer models on local data reveals distributional mismatch, enabling selective collaboration rather than forced consensus. Experimental evaluation on three wearable IoT datasets demonstrates that \textsc{Murmura} reduces performance degradation from IID to non-IID conditions by a factor of 20 compared to FedAvg (0.9\% vs 19.3\%), converges 7.4 times faster, and maintains stable accuracy across hyperparameter choices. Future work should investigate temporal dynamics where peer compatibility evolves over time, deployment on physical wearable devices to validate practical applicability, and graduated trust mechanisms that extract partial value from moderately compatible peers when few highly compatible peers exist. These results establish EDL as a principled foundation for compatibility-aware personalization in decentralized heterogeneous environments.




\section*{Acknowledgment}
This work is supported by the Australian Research Council (ARC) through Discovery Project grant DP240102088.

\bibliographystyle{IEEEtran}
\bibliography{references}

\end{document}